\documentclass[aps,twocolumn,prb,showpacs,amssymb]{revtex4}
\usepackage{epsfig}
\usepackage{bm}
\def\be{\begin{equation}}
\def\ee{\end{equation}}
\def\bea{\begin{eqnarray}}
\def\eea{\end{eqnarray}}
\def\nn{\nonumber}

\begin{document}
\title{
Boundary contributions to specific heat and susceptibility
in the spin-1/2 XXZ chain
}
\author{A. Furusaki}
\author{T. Hikihara}
\affiliation{Condensed-Matter Theory Laboratory, RIKEN,
 Wako, Saitama 351-0198, Japan}
\date{October 22, 2003}

\begin{abstract}
Exact low-temperature asymptotic behavior
of boundary contribution to specific heat and susceptibility
in the one-dimensional spin-1/2 XXZ model with
exchange anisotropy $1/2<\Delta\le1$ is analytically
obtained using the Abelian bosonization method.
The boundary spin susceptibility is divergent in the
low-temperature limit.
This singular behavior is caused by the
first-order contribution of a bulk leading irrelevant operator
to boundary free energy.
The result is confirmed by numerical
simulations of finite-size systems.
The anomalous boundary contributions in the spin isotropic case
are universal.
\end{abstract}

\maketitle

Irrelevant operators in the renormalization-group (RG) theory
are considered to give only subleading contributions to
any physical quantity by definition.
A subtle exception to this general rule can be found, however,
when a bulk irrelevant operator
has not so large boundary scaling dimension.
Consider, for example, a Gaussian theory in (1+1)-dimension
with a bulk perturbing operator $O$ having scaling dimension,
$x_o$, larger than 2.
First, let us impose periodic boundary condition to a one-dimensional
(1D) system of length $L$.
In second-order perturbation the irrelevant operator is found to give
a contribution of order $LT^{2x_o-3}$ to the specific heat
of the periodic system at temperature $T$.
This is small compared with the leading term of order $LT$.
Next, let us suppose that the system has open boundaries.
In general the specific heat has boundary contribution of order $L^0$.
Indeed, first-order perturbation leads to a boundary contribution
of order $L^0T^{x_o'-2}$ to the specific heat,
where $x_o'$ is boundary scaling
dimension of the perturbing operator.
If $2<x_o'<3$, then the boundary specific heat gives a
dominant contribution at low enough temperatures, even though
the operator is irrelevant.

This interesting possibility is in fact realized
in the 1D spin-1/2 XXZ model,
as recently pointed out by Fujimoto.\cite{Fujimoto}
The model with open boundaries was studied earlier
by de Sa and Tsvelik using Bethe ansatz,\cite{deSa}
who found anomalous boundary contributions to the specific heat
and spin susceptibility.
Similar and more extensive
calculations\cite{Essler,Asakawa,Zvyagin,FujimotoPRB}
for the spin isotropic case have shown that the boundary
contribution to the spin susceptibility at $T=0$ exhibits
a Curie-like behavior with a logarithmic correction
as a function of a weak external field.
These results seemed somewhat mysterious, as irrelevant operators
localized at a boundary can give at most a temperature-independent
contribution to the susceptibility.\cite{EggertA,FuruNaga,Frojdh}
The origin of these anomalous boundary contributions was
not understood, until Fujimoto and Eggert recently examined
boundary contributions from a bulk
leading irrelevant operator
to the specific heat and
the susceptibility.\cite{Fujimoto,FujimotoEggert}
Their theory seems rather complicated, however, as it uses properties
of boundary states in the boundary conformal field theory.
The aim of this paper is to present an alternative, simpler,
and straight-forward derivation of
the exact results using the Abelian bosonization method.
We also verify our analytical results with numerical calculations
for the anisotropic XXZ model.

Our results are relevant to many 1D correlated electron systems
with impurities.
An impurity potential in a 1D system of repulsively interacting
electrons is strongly renormalized and becomes perfectly reflecting
in the low-energy limit.\cite{Kane,FuruNaga93,Matveev}
A long wire with a finite density of impurities is thus
effectively cut into many pieces of finite length.
Measurements of the specific heat and the susceptibility of
such a wire will exhibit the anomalously large boundary contributions.

We analyze the spin-1/2 XXZ model in a weak magnetic field,
\bea
H
\!\!\!&=&\!\!\!
J\sum^{L-1}_{j=1}
\left(S^x_jS^x_{j+1}+S^y_jS^y_{j+1}+\Delta S^z_jS^z_{j+1}\right)
-h\sum^L_{j=1}S^z_j
\nn \\
&=&\!\!\!
H_0 -h \sum_{j=1}^L S^z_j,
\label{H}
\eea
where $J>0$ and $L$ is the number of sites.
We are interested in the case where the exchange anisotropy
$\Delta$ is in the parameter range
\be
\frac12 < \Delta \le1
,
\label{Delta}
\ee
for which we shall find an anomalous
contribution to the boundary free energy.

First we consider a semi-infinite spin chain,
taking the limit $L\to\infty$.
Its free energy has, in addition to a bulk term,
a boundary term of order 1 due to the
presence of the boundary at $j=1$.
We calculate the boundary free energy at low temperatures using
the low-energy effective theory\cite{Affleck}
for the Hamiltonian (\ref{H}),
which is the Gaussian model perturbed by a nonlinear operator.
The Hamiltonian density of the effective theory is
\be
\mathcal{H}=
\frac{v}{2}
\left[
\left(\frac{d\phi}{dx}\right)^2
+\Pi^2
\right]
+\lambda\cos\left(\frac{2\phi}{R}\right)
-\frac{h}{2\pi R}\frac{d\phi}{dx}.
\label{calH}
\ee
The lattice constant is unity.
The bosonic field $\phi(x)$ and its conjugate operator $\Pi(x)$,
defined on the positive half line,
satisfy $[\phi(x),\Pi(y)]=i\delta(x-y)$.
The parameter $R$ and the velocity $v$ are functions of $\Delta$
and $h$.
At zero field $h=0$, the parameter $\eta$ defined by $\eta=2\pi R^2$
is 
\be
\eta=1-\frac{1}{\pi}\cos^{-1}(\Delta),
\ee
with which the spin velocity $v$ at $h=0$ is written,
\be
v=J\frac{\sin(\pi\eta)}{2(1-\eta)}.
\ee
We note that in the parameter range (\ref{Delta}) of our interest
$\eta$ satisfies the inequality
\be
\frac{2}{3}<\eta\le 1
\label{eta-inequality}
\ee
at $h=0$.
The original spin operator can be expressed with the fields
$\phi$ and $\Pi$.
In particular, $S_j^z$ is written as
\be
S_j^z=
\frac{1}{2\pi R}\frac{d\phi}{dx}
+(-1)^{j-1}a\sin\frac{\phi(x)}{R},
\label{Sz}
\ee
where $x=j$ and $a$ is a constant calculated
in Refs.\ \onlinecite{Lukyanov99} and \onlinecite{corAmp}.
For the semi-infinite wire the bosonic field obeys the boundary
condition at $x=0$,\cite{EggertA,corAmp}
\be
\phi(x=0)=0.
\label{phi(0)}
\ee
Once Eqs.\ (\ref{Sz}) and (\ref{phi(0)}) are given, sign of
the coupling constant $\lambda$ of the nonlinear operator is fixed
to be positive.
For the gapless regime ($-1<\Delta\le1$) of the XXZ chain the exact
value of $\lambda$ is known to be\cite{Lukyanov98,LukyanovT}
\be
\frac{\lambda}{v}
=-2\sin\!\left(\frac{\pi}{\eta}\right)
\left[\frac{\Gamma(\frac{1}{\eta})}{\pi}\right]^2
\left[
\frac{\Gamma(1+\frac{\eta}{2-2\eta})}
     {2\sqrt{\pi}\,\Gamma(1+\frac{1}{2-2\eta})}
\right]^{\frac{2}{\eta}-2}
\label{lambda/v}
\ee
under the regularization condition on the zero-temperature
correlator in the Gaussian model,
\be
\langle e^{i\mu\phi(x)}e^{-i\mu\phi(y)}\rangle=|x-y|^{-\mu^2/2\pi},
\quad
x,y,|x-y|\gg1.
\label{regularization}
\ee

We are now ready to calculate the leading contribution to the
boundary free energy.
We begin with the path integral representation of the partition
function $Z$,
\be
Z=\int\mathcal{D}\phi
\exp\left(-\int^{1/T}_0d\tau\int^\infty_0dx
\mathcal{L}\right),
\ee
where the Lagrangian density in the imaginary time is
\be
\mathcal{L}=
\frac{v}{2}\left[
\left(\frac{\partial\phi}{\partial x}\right)^2
+\frac{1}{v^2}\left(\frac{\partial\phi}{\partial\tau}\right)^2
\right]
+\lambda\cos\left(\frac{2\phi}{R}\right)
-\frac{h}{2\pi R}\frac{\partial\phi}{\partial x}.
\ee
To eliminate the last term ($\propto h$), we change the field
as
\be
\phi(x,\tau)\to\phi(x,\tau)+\frac{hx}{2\pi Rv},
\label{phi-shift}
\ee
which yields a constant contribution
$-\chi_0 h^2/2$
to the bulk free energy
with the bulk susceptibility given by
$\chi_0=L/(2\pi\eta v)$.
This term is not of our main interest and will be discarded in the
following analysis.

Since the operator $\lambda\cos(2\phi/R)$ is an irrelevant
perturbation in the gapless phase, the low-temperature expansion of
the free energy can be obtained by expanding the partition function
in powers of $\lambda$.
The leading boundary contribution to the free
energy then comes from the first-order term $\propto\lambda$
in the expansion.
After making the shift (\ref{phi-shift}), the first-order term reads
\begin{widetext}
\be
Z_1=-\lambda\int\mathcal{D}\phi
\int^{1/T}_0d\tau\int^\infty_0\!dx
\cos\left(\frac{2\phi}{R}+\frac{2hx}{\eta v}\right)
\exp\left\{-\frac{v}{2}\int^{1/T}_0d\tau\int^\infty_0\!dx
\left[
\left(\frac{\partial\phi}{\partial x}\right)^2
+\frac{1}{v^2}\left(\frac{\partial\phi}{\partial\tau}\right)^2
\right]
\right\}.
\label{Z_1}
\ee
Under the boundary condition (\ref{phi(0)}) the bosonic field can be
expanded as
\be
\phi(x,\tau)=
\sqrt{\frac{2T}{\pi}}\sum_{n\in\mathbb{Z}}\int^\infty_0\!dk\,
e^{-i\omega_n\tau}\sin(kx)\tilde\phi(k,i\omega_n),
\label{mode-expansion}
\ee
where $\omega_n=2\pi n T$.
Inserting the mode expansion (\ref{mode-expansion})
into Eq.\ (\ref{Z_1}) and performing the Gaussian integral of
$\tilde\phi$, we obtain
\be
Z_1=
-Z_0\frac{\lambda}{T}\int^\infty_0\!dx\cos\left(\frac{2hx}{\eta v}\right)
\exp\!\left(-\frac{8T}{\eta v}\sum_{n\in\mathbb{Z}}\int^\infty_0\!dk
\frac{e^{-|\omega_n|/\Lambda}\sin^2(kx)}{k^2+(\omega_n/v)^2}\right),
\label{Z_1-2}
\ee
\end{widetext}
where $Z_0=Z|_{\lambda=0}$ and we have introduced the high-energy
cutoff $e^{-|\omega_n|/\Lambda}$ with $\Lambda=v$
to meet the condition (\ref{regularization}).
The integral and summation in the exponent can be easily performed,
leading to the first-order contribution to the free energy
$F=-T\ln Z$, which is nothing but the leading term in
the boundary free energy,
\be
F_1=\lambda\left(\frac{\pi T}{v}\right)^{2/\eta}
\int^\infty_\alpha dx
\frac{\cos\left(\frac{2hx}{\eta v}\right)}
     {\left[\sinh\left(\frac{2\pi Tx}{v}\right)\right]^{2/\eta}},
\label{F_1}
\ee
where $\alpha$ is a short-distance cutoff of the order of half a lattice
spacing.
Note that $F_1$ has a contribution only from the boundary region within
the distance $v/2\pi T$ from the edge of the spin chain.
The exponent $2/\eta$ is the boundary scaling dimension
of the irrelevant operator $\cos(2\phi/R)$.

Our next task is to evaluate the boundary free energy $F_1$
at low temperatures.
Let us first consider the simplest case $h=0$.
To obtain the low-temperature expansion of the integral in
Eq.\ (\ref{F_1}), we change the variable as
\be
I_1=
\int^\infty_\alpha
\frac{dx}{\left[\sinh\left(\frac{2\pi Tx}{v}\right)\right]^{2/\eta}}
=
\frac{v}{2\pi T}\int^1_{u_0}
du
u^{-\frac{2}{\eta}}(1-u^2)^{\frac{1}{\eta}-1},
\ee
where $u_0=\tanh(2\pi T\alpha/v)\approx2\pi T\alpha/v$.
For the parameter range (\ref{eta-inequality}) of our interest,
the integral is divergent in the limit $u_0\to0$.
We integrate $I_1$ by parts to separate out the diverging contribution
and obtain
\be
\frac{2\pi TI_1}{v}=
\frac{\eta}{2-\eta}u_0^{1-\frac{2}{\eta}}(1-u_0^2)^{\frac{1}{\eta}-1}
-\frac{1-\eta}{2-\eta}
B(\textstyle{\frac{3}{2}-\frac{1}{\eta},\frac{1}{\eta}-1}),
\label{I_1}
\ee
where we have set $u_0=0$ in the second term.
The first term gives regular $T^{2n}$ terms ($n=0,1,2,\cdots$)
in the free energy,
whereas the cutoff-independent second term leads to the anomalous
contribution to the boundary free energy.
From Eqs.\ (\ref{F_1}) and (\ref{I_1}) we find the leading term in
the boundary free energy
\be
F_1=E_1
-\frac{\lambda\eta}{2\sqrt{\pi}(2-\eta)}
{\textstyle\Gamma(\frac{1}{\eta})\Gamma(\frac{3}{2}-\frac{1}{\eta})}
\left(\frac{\pi T}{v}\right)^{\frac{2}{\eta}-1},
\ee
where $E_1=F_1(T=0)$ is the first-order boundary energy and
subleading terms of order $T^2$ are ignored.
Then the boundary contribution to the entropy is
\be
S_b=-\frac{\partial F_1}{\partial T}
=\frac{\sqrt{\pi}}{2}
{\textstyle\Gamma(\frac{1}{\eta})\Gamma(\frac{3}{2}-\frac{1}{\eta})}
\frac{\lambda}{v}
\left(\frac{\pi T}{v}\right)^{\frac{2}{\eta}-2},
\label{S}
\ee
and the boundary contribution to the specific heat is
\be
C_b=T\frac{\partial S_b}{\partial T}
=\sqrt{\pi}
{\textstyle
(\frac{1}{\eta}-1)\Gamma(\frac{1}{\eta})\Gamma(\frac{3}{2}-\frac{1}{\eta})
}
\frac{\lambda}{v}\left(\frac{\pi T}{v}\right)^{\frac{2}{\eta}-2}.
\label{C}
\ee
For the XXZ chain with $1/2<\Delta<1$, Eqs.\ (\ref{C}) and
(\ref{lambda/v}) give the exact leading boundary
contribution to the specific heat.
For $\Delta<1/2$ the leading term in the boundary specific heat
is proportional to $T$.
At $\Delta=1/2$ there appears a logarithmic correction to the $T$-linear
specific heat.

For a small but finite $h$ we may expand $\cos(2hx/\eta v)$ in
Eq.\ (\ref{F_1}) as
\be
\int^\infty_\alpha\!\!dx
\frac{\cos\left(\frac{2hx}{\eta v}\right)-1}
     {\left[\sinh\left(\frac{2\pi Tx}{v}\right)\right]^{2/\eta}}
\approx
-\frac{2h^2}{\eta^2v^2}\int^\infty_0 \!
\frac{x^2dx}{\left[\sinh\left(\frac{2\pi Tx}{v}\right)\right]^{2/\eta}},
\ee
where we have taken the limit $\alpha\to0$ in the convergent integral.
The integral in the right-hand side is elementary.
After some algebra we find
\be
\int^\infty_0\!\frac{x^2dx}{(\sinh x)^{2/\eta}}
=\frac{2^{\frac{2}{\eta}-4}\eta \Gamma(\frac{1}{\eta})\Gamma(3-\frac{2}{\eta})}
      {(2-\eta)\Gamma(2-\frac{1}{\eta})}
[\pi^2-2\psi'({\textstyle\frac{1}{\eta}})],
\ee
where $\psi'(x)=d^2\ln\Gamma(x)/dx^2$.
The boundary contribution to the susceptibility is then obtain as
\bea
\chi_b
\!\!\!&=&\!\!\!
-\left.\frac{\partial^2F_1}{\partial h^2}\right|_{h=0}
\nn\\
&=&\!\!\!
\frac{\lambda\Gamma(\frac{1}{\eta})\Gamma(3-\frac{2}{\eta})
      [\pi^2-2\psi'({\textstyle\frac{1}{\eta}})]}
     {4v^2\eta(2-\eta)\Gamma(2-\frac{1}{\eta})}
\left(\frac{2\pi T}{v}\right)^{\frac{2}{\eta}-3}.
\label{chi}
\eea
This is the exact leading contribution to the boundary
susceptibility in the XXZ chain with $1/2<\Delta<1$.
Note that $\chi$ is diverging as $T\to0$.
For $\Delta<1/2$ the leading term in the susceptibility is
independent of temperature.
A logarithmic correction shows up at $\Delta=1/2$.

At $T=0$ the boundary energy $E_1$ becomes
\bea
E_1
\!\!\!&=&\!\!\!
\lambda\int^\infty_\alpha\!dx
\frac{\cos\left(\frac{2hx}{\eta v}\right)}{(2x)^{2/\eta}}
\nn\\
&=&\!\!\!
\frac{\lambda(2\alpha)^{1-\frac{2}{\eta}}}{2(\frac{2}{\eta}-1)}
\cos\!\left(\frac{2h\alpha}{\eta v}\right)
\nn\\
&&
-\frac{\lambda\Gamma(3-\frac{2}{\eta})\sin[\pi(\frac{1}{\eta}-1)]}
      {4(\frac{2}{\eta}-1)(\frac{1}{\eta}-1)}
\left(\frac{h}{\eta v}\right)^{\frac{2}{\eta}-1},
\label{E_1}
\eea
from which we obtain the boundary contribution to
the total magnetization
\be
M_b=-\frac{\partial E_1}{\partial h}
=\frac{\pi\lambda}{4\eta v}
\frac{\Gamma(3-\frac{2}{\eta})}
     {\Gamma(\frac{1}{\eta})\Gamma(2-\frac{1}{\eta})}
\left(\frac{h}{\eta v}\right)^{\frac{2}{\eta}-2},
\label{M}
\ee
where higher-order terms $\mathcal{O}(h)$ are ignored.
This is the exact form of the leading term in the XXZ chain
for $1/2<\Delta<1$.
The leading terms in $C_b(T)$, $\chi_b(T)$, and $M_b(h)$ we
calculated above are found in agreement with those obtained
independently by Fujimoto and Eggert,\cite{FujimotoEggert}
after some analytical transformations are made in their results.

At the isotropic antiferromagnetic Heisenberg point ($\Delta=1$), 
we cannot substitute $\lambda$ of Eq.\ (\ref{lambda/v})
into Eqs.\ (\ref{C}), (\ref{chi}), and (\ref{M}), as it vanishes
at $\eta=1$.
To circumvent this difficulty, we adapt a RG
improved perturbation theory in which we replace a constant
$\lambda$ with a running coupling constant $\lambda(E)$
at energy scale $E$.
At $h=0$, the one-loop scaling equations
for the sine-Gordon Hamiltonian (\ref{calH}) are given by
\be
\frac{dg}{dl}=\left(2-\frac{1}{\pi R^2}\right)g,
\qquad
\frac{dR}{dl}=\frac{\pi g^2}{2R},
\label{RG}
\ee
where $g=\lambda/v$ and $dl=-d\ln E$.
The coupling $g$ is marginally irrelevant at $R=1/\sqrt{2\pi}$
(i.e., at $\Delta=1$) and decreases to zero logarithmically.
To see this explicitly, we may write $\sqrt{2\pi}R=1-\epsilon$
and expand the RG equations (\ref{RG}) in lowest order in $\epsilon$.
We solve the simplified RG equations to find the solution
corresponding to the antiferromagnetic Heisenberg chain,
\be
\frac{\lambda(l)}{v}=\frac{1}{2\pi l},\qquad
R(l)=\frac{1}{\sqrt{2\pi}}\left(1-\frac{1}{4l}\right)
\label{solution}
\ee
for $l\gg1$.
To obtain the leading term in $C_b(T)$, $\chi_b(T)$, and $M_b(h)$,
we may substitute $\eta=1$ and
\be
\lambda(T,h)=
\frac{v}{\displaystyle2\pi\ln\left(\frac{1}{\max(T,h)}\right)}
\ee
into the analytic formulas of the boundary free energy
we obtained for $1/2<\Delta<1$.
For example, the boundary entropy $S_b$ is found from Eq.\ (\ref{S})
to be
\be
S_b=\frac{\pi\lambda(T)}{2v}=\frac{1}{4\ln(T_0/T)},
\ee
from which the boundary specific heat becomes
\be
C_b=\frac{1}{4[\ln(T_0/T)]^2},
\label{isotropic-C}
\ee
in agreement with Ref.\ \onlinecite{FujimotoEggert}.
Here $T_0$ is a constant of order $J$.
The boundary contribution to the susceptibility is obtained
from Eq.\ (\ref{chi}) as
\be
\chi_b=\frac{\lambda(T)}{8\pi vT}[\pi^2-2\psi'(1)]
=\frac{1}{24T\ln(T_0/T)}.
\label{isotropic-chi}
\ee
This result was confirmed later by the corrected
calculation in Ref.~\onlinecite{FujimotoEggert}.
The boundary contribution to the magnetization at zero temperature
is found from Eq.\ (\ref{M}) to be
\be
M_b=\frac{\pi\lambda(h)}{4v}=\frac{1}{8\ln(h_0/h)},
\label{isotropic-M}
\ee
in agreement with the exact result from the Bethe
ansatz.\cite{Essler,Asakawa,Zvyagin}
Here $h_0$ is another constant of order $J$.
We note that there exist subleading terms to these formulas which are
reduced only by a factor of $1/\ln(T_0/T)$ or $1/\ln(h_0/h)$.
Even though they are important for quantitative analysis, we do not
try to evaluate it here, as it is beyond the scope of this paper.

Finally, we present comparison between the analytic result of
Eq.\ (\ref{E_1}) and numerical data of energy eigenvalues
of the XXZ spin chains with open boundaries.
Using the density-matrix renormalization-group (DMRG) 
method,\cite{White1,White2}
we have computed the lowest energy $E(L,m)$ of $H_0$ in 
the subspace of the magnetization per site
$m = (1/L) \sum_{j=1}^L S^z_j$.
The maximum system size we calculated is $L = 800$.
The number of block states kept in the DMRG calculation
is up to 200.
The numerical errors due to the DMRG truncation, 
estimated from the difference between the data 
with 150 and 200 states kept, are typically less than $10^{-5}$.
The numerical data are therefore accurate enough 
for the discussion below.

\begin{figure}
\begin{center}
\noindent
\epsfxsize=0.45\textwidth
\epsfbox{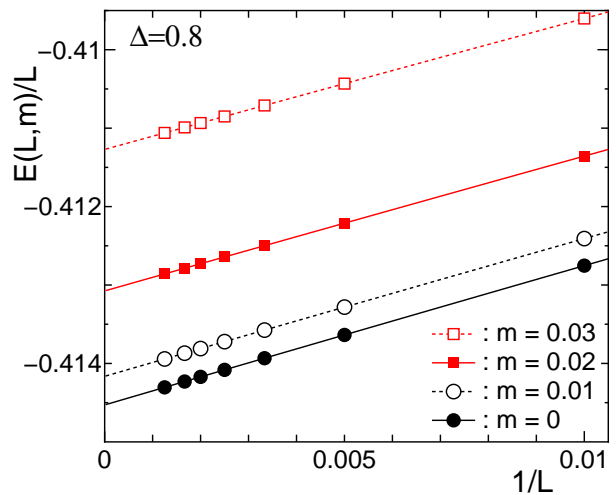}
\end{center}
\caption{
$E(L,m)/L$ as functions of $1/L$ for $\Delta = 0.8$.
The curves represent the fitting to Eq.\ (\ref{eq:EL}).
}
\label{fig:EL}
\end{figure}

\begin{figure}
\begin{center}
\noindent
\epsfxsize=0.45\textwidth
\epsfbox{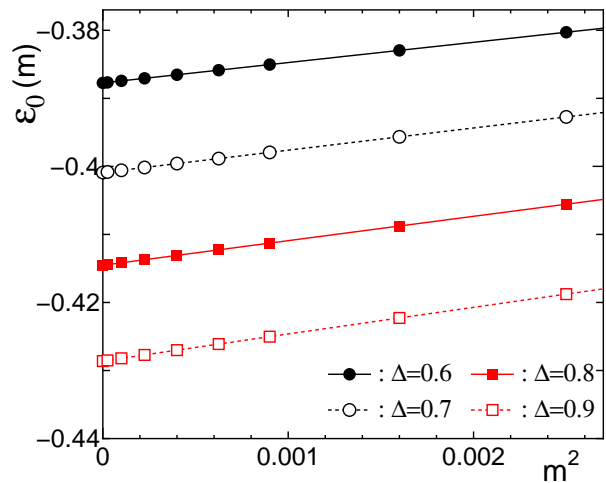}
\end{center}
\caption{
Numerical estimates of $\varepsilon_0(m)$ (symbols) and 
their exact values (curves) as functions of $m$ 
for several typical values of $\Delta$.
}
\label{fig:eps0}
\end{figure}

The energy $E(L,m)$ is known to depend on the system size $L$ as
\be
E(L,m)=L\varepsilon_0(m)+\varepsilon_1(m)+\frac{\varepsilon_2(m)}{L+1}.
\label{eq:EL}
\ee
The bulk energy $\varepsilon_0(m)$
can be obtained exactly by solving
Bethe ansatz integral equations,\cite{BogoIK,QinFYOA,CabraHP} 
while the boundary energy $\varepsilon_1(m)$ is related to
Eq.\ (\ref{E_1}).
Here, we fit the numerical data to the formula (\ref{eq:EL})
taking $\varepsilon_0$, $\varepsilon_1$, and $\varepsilon_2$ 
as free parameters.
The data of $E(L,m)$ at $\Delta = 0.8$ are shown in Fig.\ \ref{fig:EL} 
for several values of $m$.
We have found that the fitting works pretty well for
all $\Delta$s and $m$s calculated.
In fact, the estimated values of $\varepsilon_0$ obtained from the
fitting are in excellent agreement with the exact values calculated
from the Bethe ansatz equations (see Fig.\ \ref{fig:eps0}).
The bulk energy $\varepsilon_0(m)$ has smooth quadratic dependence on
$m$ since the irrelevant operator $\cos(2\pi/R)$ yields only subleading
contributions to the bulk quantity.

\begin{figure}
\begin{center}
\noindent
\epsfxsize=0.45\textwidth
\epsfbox{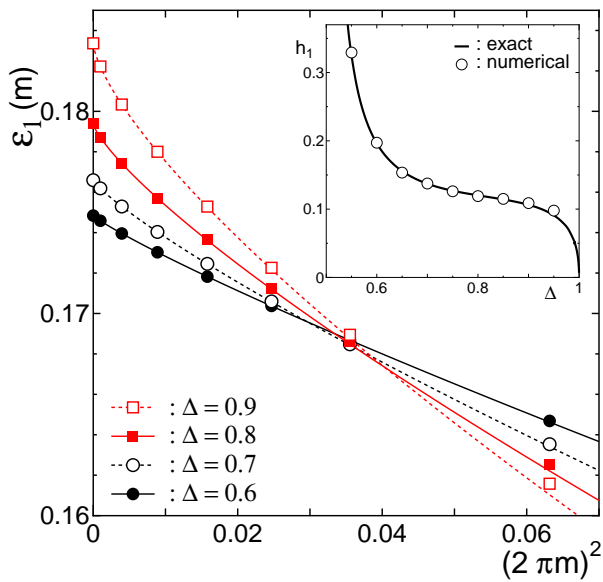}
\end{center}
\caption{
$\varepsilon_1(m)$ as functions of $(2 \pi m)^2$ 
for typical values of $\Delta$. 
The curves are the fitting to Eq.\ (\ref{fitting}).
Inset: Numerical estimates of $h_1$ as a function of $\Delta$
(open circles) and the exact result, Eq.\ (\ref{h_1}) (solid curve). 
}
\label{fig:eps1}
\end{figure}

According to Eq.\ (\ref{E_1}) and the relation between the bulk
magnetization and the external field ($m=\chi_0 h/L$),
the boundary energy $\varepsilon_1(m)$ 
should depend on $m$ as
\be
\varepsilon_1(m)=\varepsilon_1(0)-h_1 (2\pi m)^{\frac{2}{\eta}-1}
-\frac{(2\pi m)^2}{2\chi_2}
\label{fitting}
\ee
for small $m$,
where $\chi_2$ is a constant and the exact form of $h_1$ is
\be
h_1=\frac{\lambda\Gamma(3-\frac{2}{\eta})\sin[\pi(\frac{1}{\eta}-1)]}
         {2(\frac{2}{\eta}-1)(\frac{1}{\eta}-1)}.
\label{h_1}
\ee
Note that the boundary energy of the open spin chains should be twice
that of the semi-infinite wire that we have considered
in the analytic calculation.
To estimate the coefficient $h_1$ numerically, 
we fit the data of $\varepsilon_1(m)$ to Eq.\ (\ref{fitting}) 
taking $\varepsilon_1(0)$, $h_1$, and $\chi_2$ as free parameters.
We show the $m$ dependence of $\varepsilon_1(m)$
and the fitting curves in Fig.\ \ref{fig:eps1}.
Clearly, the plot of $\varepsilon_1(m)$ versus $(2\pi m)^2$ shows
large deviation from linear dependence on $(2\pi m)^2$,
confirming the existence of the leading power-law term with the
exponent $2/\eta-1$, 
which emerges from the irrelevant operator $\lambda \cos(2\phi/R)$.
The values of $h_1$ estimated from the fitting procedure are also 
shown in Fig.\ \ref{fig:eps1}. 
We see that the numerical estimates agree extremely well 
with the analytic result (\ref{h_1}).
This agreement can be regarded as a numerical proof 
of our analytic result, Eq.\ (\ref{E_1}).

In summary, we have obtained the exact leading boundary contributions
to the specific heat and the susceptibility using the low-energy
effective theory for the spin-1/2 XXZ model.
Our method can be easily extended to other 1D systems
(e.g., the extended Hubbard model and spin ladders in high
magnetic fields)
as long as their low-energy effective theory is given by
Eq.\ (\ref{calH}),
which contains nonuniversal constants $v$ and $\lambda$ to be
determined with some other method, however.
Our formulas (\ref{isotropic-C}), (\ref{isotropic-chi}), and
(\ref{isotropic-M}) for the spin isotropic case ($\Delta=1$)
are universal,
in the sense that if we make any moderate deformation of the isotropic
antiferromagnetic Heisenberg model that respects the SU(2) and
translational symmetry and that does not generate a spin gap,
then these formulas apply to some values of $T_0$ and $h_0$, 
except for cases where the coupling $\lambda$ is fine tuned
to vanish.
This implies that, for example, the 1D (extended) Hubbard model
and other systems of interacting electrons with the spin SU(2)
symmetry should universally exhibit the boundary contributions
(\ref{isotropic-C}), (\ref{isotropic-chi}), and
(\ref{isotropic-M}).

\acknowledgments
We are grateful to S.\ Fujimoto for sharing his results prior
to publication.
This work was supported in part by a Grant-in-Aid for
Scientific Research on Priority Areas (Grant No.\ 12046238)
and by NAREGI Nanoscience Project, Ministry of
Education, Culture, Sports, Science and Technology, Japan.

\end{document}